\documentstyle[12pt,aasms4]{article}

\begin{document}

\title{Intermittent behavior of cosmic mass field revealed by QSO's
Ly$\alpha$ forests}

\author{Priya Jamkhedkar\altaffilmark{1}, Hu Zhan\altaffilmark{2},
  and Li-Zhi Fang\altaffilmark{1}}

\altaffiltext{1}{Department of Physics, University of Arizona, Tucson,
AZ 85721}

\altaffiltext{2}{Department of Physics and Astronomy, Arizona State
University, Tempe, AZ 85287}

\begin{abstract}

The intermittent behavior of the space-scale distribution of Ly$\alpha$
transmitted flux of QSO HS1700+64 has been analyzed via a discrete
wavelet transform. We found that there are strong indications of
intermittency on scales down to about 10 $h^{-1}$ kpc. These are: 1.) the
probability distribution function of the local fluctuations of the flux
is significantly long-tailed on small scales, and 2.) the local power
spectrum of the flux shows prominent spiky structures on small scales.
Moreover, the local power spectrum averaged on regions with different sizes
shows similar spiky structures. Therefore, the random mass density field 
traced by the Ly$\alpha$ forests is rougher on smaller scales, consistent 
with singular clustering.

\end{abstract}


\section{Introduction}

The structures of the cosmic mass field on scales from sub-Mpc to kpc
have attracted a lot of attention recently. High resolution N-body
simulations show that the core profiles of massive halos of the cold
dark matter (CDM) cosmogony are singular (Navarro, Frenk \& White
1997, Moore et al. 2000, Jing \& Suto 2000), while the halo profiles
required by the rotation curves of dwarf galaxies (Flores \& Primack
1994, Burkeret 1995) are shallower than the numerical results.
The central cusps of dark halos also disagree with soft halo
profiles inferred from low surface brightness galaxies (de Blok \&
McGaugh 1997.) The cores of galaxies and clusters are even found to be
consistent with the thermal equilibrium model with an ``universal''
mass density (Firmani et al. 2000.)  In other words, the singular
behavior of cosmic clustering has not been detected in the cores
of galaxies and clusters.

We can now ask the question: can the singular behavior, if it exists, be
revealed by methods other than the mass profile of galaxies and
clusters? At the first glance, this goal seems to be not achievable,
as singular mass density profiles can only be seen in cores of galaxies
and clusters. However, a random mass field $\rho(x)$ consisting of rare
singular structures randomly scattered in low mass density background
typically is intermittent (Zel'dovich et al. 1990.) A basic character
of an intermittent field is that the density {\it difference} between two
neighboring positions, $|\rho(x+r)-\rho(x)|$, can be ``abnormally''
large when $r$ is very small. That is, the rare events of large density
difference $|\rho(x+r)-\rho(x)|$ on small scales $r$ have a higher
probability than that for a Gaussian field. Such a singular behavior of
a random mass field means that the probability density function (PDF)
of the density differences on small scales, $r$, has a long tail.
Obviously, the effects of long tailed PDF of the density difference are
not limited to the singular mass profile. This motivated us to look for
the PDF's long tail and its effects by using samples other than the
cores of galaxies and clusters.

The PDF's long tail have not yet been seriously studied. The most popular
statistical measure of large-scale structures -- the power spectrum of
a mass field, is insensitive to the PDF's long tail. Furthermore, The
density difference, $|\rho(x+r)-\rho(x)|$, is a quantity localized in
space $x$ and on scale $r$, and so a space-scale decomposition is
necessary. Thus using the power spectrum or any statistic that is not
based on proper space-scale decomposition, it is not possible to identify
the effects of long-tailed PDF of the density difference.

In this {\it Letter}, using a discrete wavelet transform (DWT), we look
for the long tail effects from a sample of QSO's Ly$\alpha$ forests. It
is believed that the distribution of baryonic diffuse matter is almost
point-by-point proportional to the underlying dark matter density.
Moreover, the absorption optical depth of Ly$\alpha$ is linearly
dependent on the baryonic density. Therefore, the high resolution data
of the transmitted flux of QSO's absorption would be a good candidate to
reveal the long-tailed PDF of cosmic mass field on small scales.

\section{Method}

Let us consider a 1-D random mass density field $\rho(x)$ in spatial
range $L$. With a DWT space-scale decomposition, the local density
difference, $|\rho(x+r)-\rho(x)|$, is represented by the wavelet function
coefficients (WFCs) as
\begin{equation}
\tilde{\epsilon}_{j,l} =\langle \psi_{j,l}, \rho \rangle,
\end{equation}
where $\psi_{j,l}(x)$ is the orthonormal and complete basis of the
discrete wavelet transform, and $\langle ... \rangle =\int ....dx$ is
the inner product (Daubechies, 1992.) We use DAUB4 wavelet (Press et al.
1993, Nielsen 1998) for our analysis through out this {\it Letter}. The WFC,
$\tilde{\epsilon}_{j,l}$, is the density fluctuation on the scale $L/2^j$
at the position $l=0,...2^{j}-1$, or the  mean density difference between
nearest neighbors ranging on scale $L/2^j$ at $l$. If the ``fair sample
hypothesis" (Peebles 1980) holds, then the $2^j$ values of
$\tilde{\epsilon}_{j,l}$ form an ensemble of the density differences on
scale $j$, and therefore,
the distribution of $\tilde{\epsilon}_{j,l}$ is a reasonable estimate
of the PDF of the  density differences on scale $j$ (Fang \& Thews 1998.)

The second order statistics $|\tilde{\epsilon}_{j,l}|^2$ describes
the power of the perturbations of the mode $(j,l)$. In other words,
at a given position $l$, the local power spectrum is given by
\begin{equation}
P_{j,l} = \tilde{\epsilon}_{j,l}^2.
\end{equation}
By averaging $P_{j,l}$ over all positions  $l$, we have
\begin{equation}
P_j =\frac{1}{2^j}\sum_{l=0}^{2^j-1}
 |\tilde{\epsilon}_{j,l}|^2 = \frac{1}{2^j}\sum_{l=0}^{2^j-1}P_{j,l}.
\end{equation}
It has been shown that $P_j$ actually is a band-averaged Fourier
power spectrum (Pando \& Fang 1998; Fang \& Feng 2000.)

The Fourier power spectrum lacks phase information, and therefore,
$P_j$ cannot show the phase-related features of clustering. However,
$P_{j,l}$ is phase-sensitive. One can search for the phase-related
features of the mass field by {\it local} DWT power spectrum $P_{j,l}$.

We can generalize the definition of local DWT power spectrum, eq.(2)
as follows. First we chop $L$ into $2^{j_s}$ sub-interval, labeled by
$l_s= 0, 1...(2^{j_s}-1)$. Each sub-interval has a length $L/2^{j_s}$.
Then, the local DWT power spectrum at sub-interval $l_s$ is given by
\begin{equation}
P_{j,\{j_s,l_s\}}=\frac{1}{2^{j-j_s}}
 \sum_{l=l_s2^{j-j_{s}}}^{(l_s+1)2^{j-j_{s}}-1}
 |\tilde{\epsilon}_{j,l}|^2.
\end{equation}
It is the power on scale $L/2^j$ localized on $l_s$ with size
$L/2^{j_s}$.

For a Gaussian field, the local power spectrum $P_{j,l}$ will not
show structures with respect to $l$. On the other hand, the singular
behavior of a random field is measured by the exponent $\alpha$
defined by $|\rho(x+r)-\rho(x)| \sim r^{\alpha}$. The larger the
$\alpha$ is, the smoother the field on small scales is, and vice
versa. If the exponent $\alpha$ is negative there are an actual
singularity of the field. Therefore, the singular behavior can be
revealed by the roughness of the local power spectrum on small 
scales. The WFC local power spectrum can also measure the index 
$n$ of power-law profile $\rho \sim r^{-n}$ for individual core.

\section{Sample and analysis}

The sample used for the analysis is the Ly$\alpha$ transmitted
flux of QSO HS1700+64. This sample has been employed to study the evolution
of structure (Bi \& Davidsen 1997), the Fourier and DWT power spectra
(Feng \& Fang 2000.) The recovered power spectrum has been found
to be consistent with the CDM model on scales larger than about 0.1
$h^{-1}$ Mpc. The data ranges from 3727.012\AA \  to 5523.554\AA \  with a
resolution of $3$ kms$^{-1}$, for a total of 55882 pixels. In this paper,
we use the first 25000 pixels for analysis, which correspond to $z =$
2.07$\sim$ 2.65, or $\lambda =$ 3727.012\AA \ $\sim$ \ 4434.266\AA. On
average, a pixel is about 0.029\AA, equivalent to physical size $\sim 5$
$h^{-1}$ kpc at $z \sim 2$ for a flat universe. We pad
7768 null pixels at the end to utilize a fast wavelet transform algorithm,
which requires the data size in powers of 2. It does not affect the
analysis,
because the wavelet transform is localized. Moreover, we subject DWT
directly to pixels without transforming them to physical positions.
The relation between pixel number and physical position is not linear,
but it does not affect structures on small scales. Thus, we ignore the
effect of the non-linear relation in our present analysis.

Most lines in the Ly$\alpha$ transmitted flux are due to absorptions
by gases in cool and low density regions. The pressure gradients are
generally less than gravitational forces. That is, the gas, and hence
the transmitted flux, should be good tracer of the dark matter.
Nevertheless, small scale structures of the dark matter field may be
smoothed out by the velocity dispersion of Ly$\alpha$ forest gases.
Therefore, to identify the clustering feature, we will statistically
compare the real data with its phase-randomized counterpart,
which is obtained by taking the inverse transform of the Fourier
coefficients of the original data after randomizing their phases uniformly
over $[0,2\pi]$ without changing their amplitudes.

\subsection{The PDFs of WFCs}

In Fig. 1, we show the PDFs of the WFCs for $j=8$, and 14. Each PDF is
normalized to have unit variance. For scale $j=8$, the departure from
Gaussian distribution is not so significant. Especially, no tail shows
in the $j=8$ PDF, i.e. no WFCs found to be
$\tilde{\epsilon}_{j,l} \geq 3 \sigma$. For j = 14, the PDF of WFCs has
$2^{14}$ events. Therefore, if the PDF were Gaussian, the number of
the events larger than 3$\sigma$ would be about 44. However, the data
shows 234 events beyond the 3$\sigma$ range. Furthermore, a Gaussian PDF
predicts that the number of the events larger than 5$\sigma$ should be 0.01,
while the data shows more than 100 events larger than 5$\sigma$. The data
extends to beyond 15$\sigma$ on both sides. Therefore, the PDF is
indeed significantly long-tailed on small scales. In other words, the
field is rougher on smaller scales. This indicates that the field may
contain singular structures.

The shape of the two PDFs of $j=8$ and 14 is very different from each other.
That is, the two stochastic variables $\tilde{\epsilon}_{j,l}$, $j=8,
\ 14$ don't relate to each other as
\begin{equation}
 \tilde{\epsilon}_{j,l}
  = 2^{\beta (j-j')}
   \tilde{\epsilon}_{j',l},
\end{equation}
where $\beta$ is a constant. Therefore, the mass field traced by the
QSO HS1700+64 is unlikely to be self-similar.

\subsection{Local power spectrum}

In Fig. 2 we plot the local DWT power spectra of the HS1700+64
transmitted flux and its phase-randomized counterpart.
We take $j_s=7$, i.e. chopping the entire sample into 128 sub-intervals,
and, in each sub-interval, calculating the power spectra for
$j= 8$ $\sim$ 14, which correspond to physical scales
$2^{15-j} \times 5$ $h^{-1}$ kpc.

Fig. 2 shows that the $j=8$ local power spectrum for real data is
not very different from its phase-randomized counterpart. This is
consistent with the $j=8$  PDF shown in Fig.1. It is closer-to-Gaussian.
While the $j=12$ $\sim$  $14$ local spectra are very rough, showing
remarkably spiky structures, that completely disappear in the phase
randomized counterpart. The spiky features mean that a significant
part of the power is concentrated in some small areas. This feature is
a result of the long tailed PDF of the density difference, i.e., higher
probability of ``abnormal'' density change. The smaller the scale, the
more pronounced the spiky features. This, again, points to a singular
clustering of cosmic mass field.

It should be emphasized that the spikes in the local power spectrum
with high $j$ do {\it not} always correspond to the peaks in the density
distribution (or the absorption lines in the optical spectrum). The WFCs
$\tilde{\epsilon}_{j,l}$ describe the {\it difference} in density between
intervals of length $2^{15-j}\times 5$ $h^{-1}$ Mpc. The average of
$\tilde{\epsilon}_{j,l}$ over $l$ generally is zero. The mean power
(or variance) at $j=14$ is $P_{14}= 1.2\times10^{-5}$. Thus, even a single event 
$>10\sigma$ at $j=14$ doesn't always refer to high density, and it can
happen in regions other than high density cores. The spikes denote the
positions where the density (or the absorption optical depth) undergoes
a dramatic change, which is the key indicator of the singular behavior of
a random field.

The wavelet functions $\psi_{j,l}$ are orthogonal, and
therefore, the local power spectrum on scale $j$ doesn't contaminate
perturbations on other scales. This is very different from
the density distribution smoothed by a window function on scale $j$.
The peaks identified from the window-function smoothed density
field contain all contributions from perturbations on scales
$j' \leq j$. Therefore, the peaks in a smoothed field are
actually given by a superposition of perturbations on large and small
scales. They may not show singular features, because the PDFs of
large scale (or $j' < j$) perturbations are closer to Gaussian.

We should estimate the possible distortion of the long tail effects
caused by velocity dispersion of gases. We calculated
the $j=14$ local power spectrum with sub-interval $j_s=14$, which is
plotted in Fig. 3. This local power spectrum has almost the same spiky
features identified in the $j=14,\ j_s = 7$ spectrum (see, Fig. 2.)
That is, most spikes shown by the power localized in sub-interval with size
of about 600 $h^{-1}$ kpc actually are localized in sub-interval
with size only about few 10 $h^{-1}$ kpc. This result indicates that the
contamination of gas velocity dispersion may not be significant,
at least, for prominent spikes. Thus, the spiky structures should mainly
come from the underlying mass field.

Intermittency can more clearly be seen in Fig. 3. The mean powers
[eq.(3)] of the real data (left panel) and randomized counterpart (right
panel) of the $j_s=14$ and $j=14$ local spectrum actually are the same,
while the spikes of the real data are higher than the mean power by a
factor of few tens even hundreds. That is, in the real sample, most power
of the $j=14$ perturbations is concentrated in the spikes, and almost no
power, i.e. $P_{j,\{j_s,l_s\}}\simeq 0$, in places other than the spikes.
This is a typical intermittent distribution.

\section{Conclusions}

With the PDFs of $\tilde{\epsilon}_{j,l}$ and local DWT power spectra,
we show that the mass field traced by the QSO HS1700+64 Ly$\alpha$
forests is neither Gaussian, nor self-similar, but intermittent. The
spiky features shown in the local DWT power spectrum is remarkably
pronounced on scales down to about 10 $h^{-1}$ kpc. Moreover, the long
tail and the spiky features are substantial on smaller scales. This
indicates that the cosmic mass field is rougher on smaller scales,
consistent with singular clustering.

A big advantage of intermittency is that one can detect singular
clustering using the statistical features of entire random density
field, not limited to the cores of galaxies and clusters. The information
of intermittency extracted from Ly$\alpha$ forests would be important
to test models of cosmic clustering in terms of their singular behavior.

\acknowledgments

We thank Dr. D. Tytler for kindly providing the data of the Keck spectrum
HS1700+64. PJ would also like to thank Dr. Robert Maier for his help. HZ
thanks Dr. David Burstein for helpful discussions.

\clearpage
\begin{figure}
\newpage
\epsscale{0.8}
\plotone{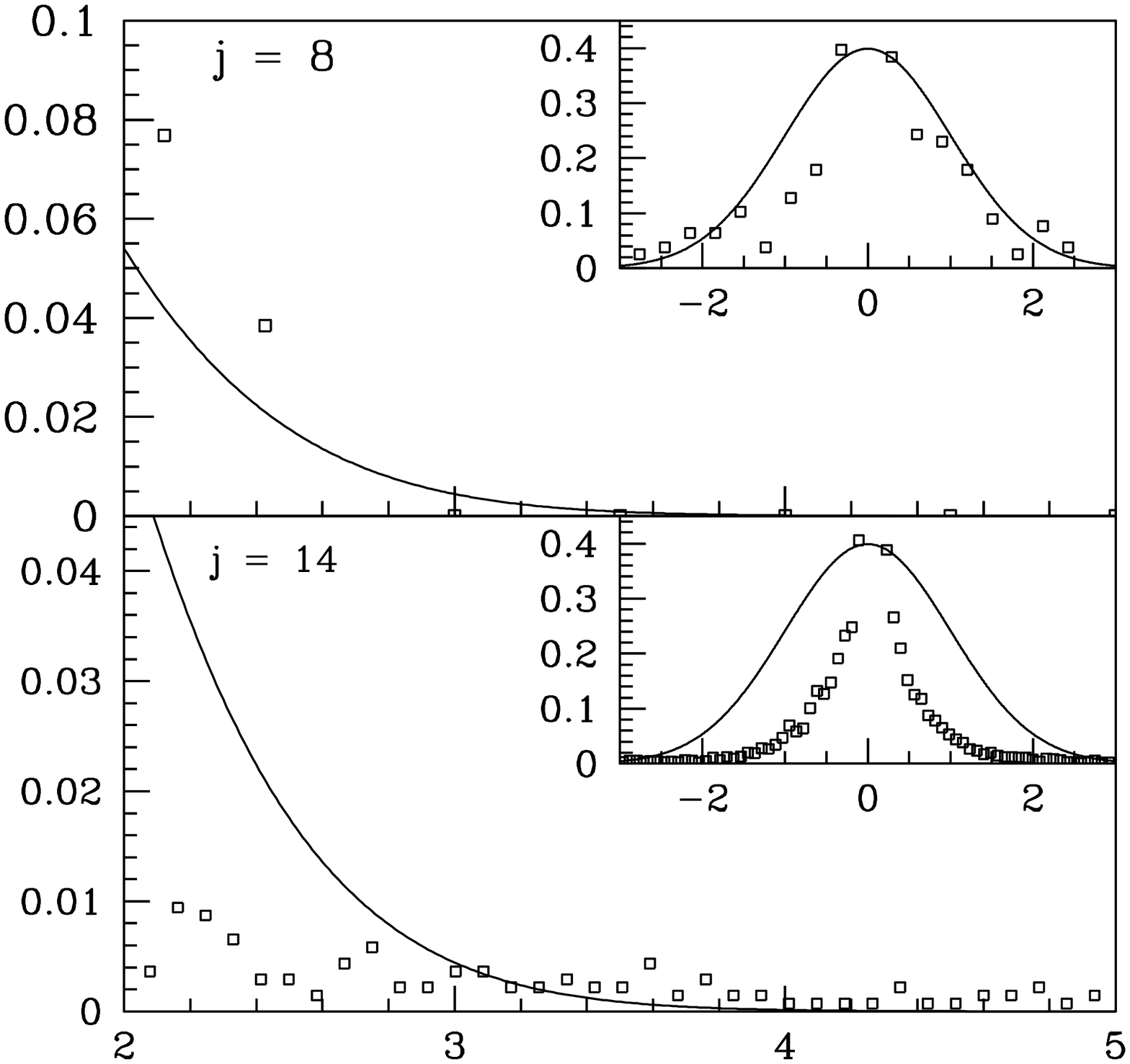}
\figcaption[fig1.ps] {The PDFs of the WFCs $\tilde{\epsilon}_{j,l}$
for $j=8$ and $j=14$ are shown on top right corner of each panel.
The tail is shown on the main part each panels. Horizontal axis is
for $\tilde{\epsilon}_{j,l}/\sigma$, where $\sigma$ is the variance
of the sample. Vertical axis is the probability density. The solid
curve is the Gaussian distribution with zero mean and unit variance.
The PDF of $j=14$ at zero is $\sim$4.5, so it is not shown in the
figures.
\label{fig1}}
\end{figure}

\clearpage
\begin{figure}
\newpage
\epsscale{0.9}
\plotone{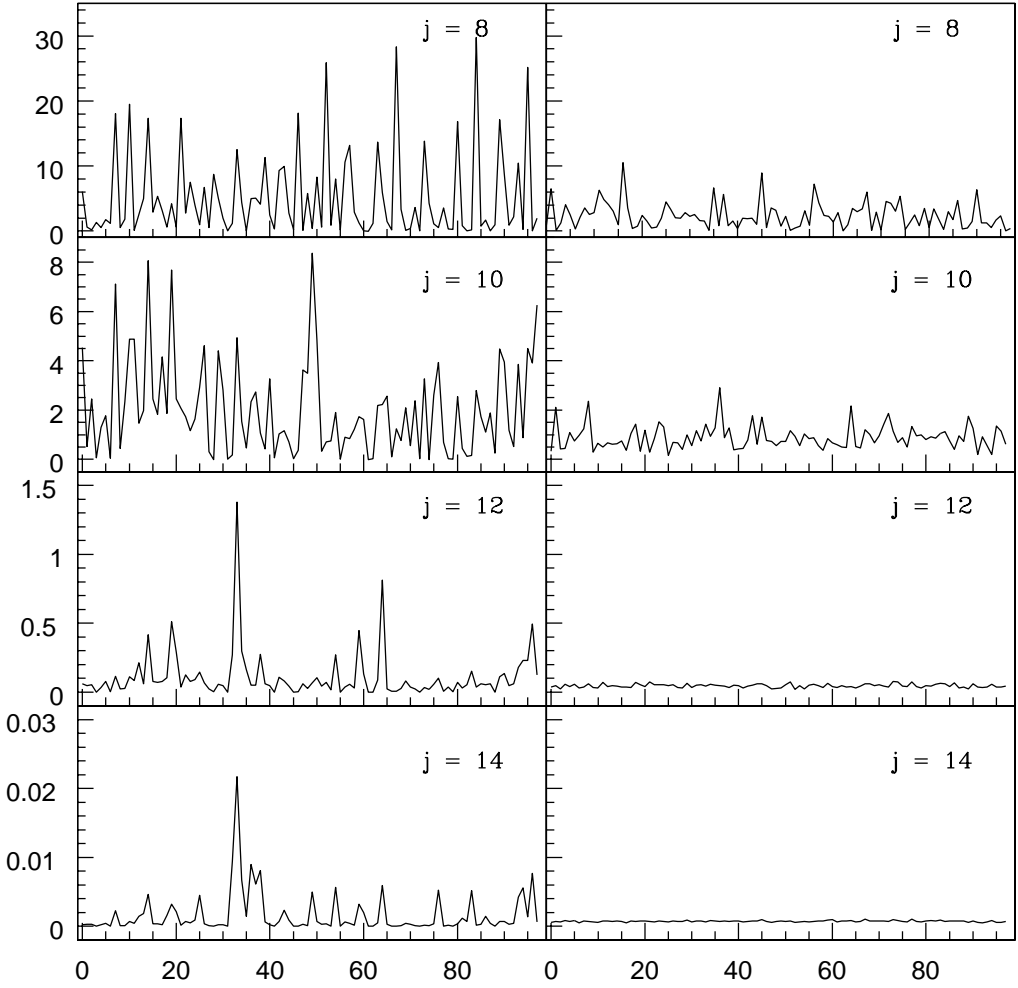}

\figcaption[fig2.ps] {The vertical axis represents
$2^{j-j_s}P_{j,\{j_s,l_s\}}$ with $j_s=7$. The left and right panels
are for the HS1700+64 transmitted flux, and its phase-randomized
counterpart, respectively. The horizontal axis represents the position
$l_s$ in units of $2^7 \times 10$ $h^{-1}$ kpc. The scales of the local
power spectra, $j= 8,\ 10,\ 12,\ 14$, correspond to physical scales
$2^{15-j} \times 5$ $h^{-1}$ kpc.
\label{fig2}}

\end{figure}

\clearpage
\begin{figure}
\newpage
\epsscale{0.8}
\plotone{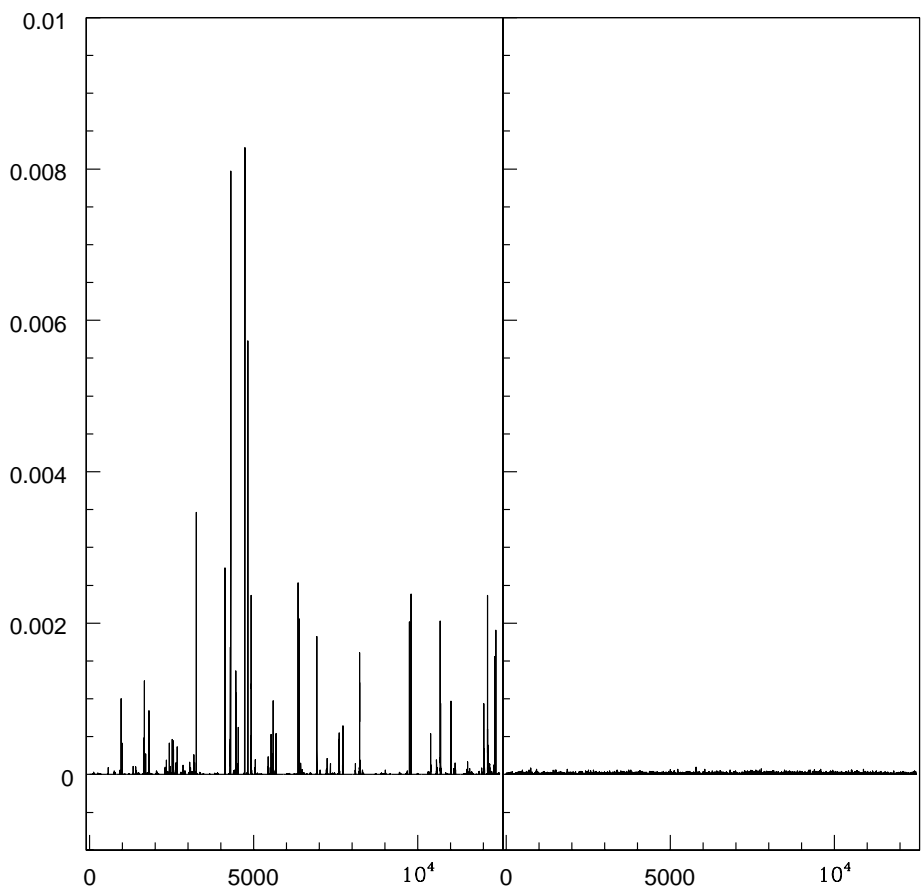}
\figcaption[fig3.ps]{The same as Fig. 2 with $j=14$, but taking
$j_s=14$.
 \label{fig3}}
\end{figure}

\end{document}